# Analysis of Jet Dynamics and Collimation Characteristics of 0241+622 on Parsec Scales

Haitian Shang[1,2,3], Wei Zhao[1], Xiaoyu Hong[1,2,3,4], and Xu-zhi Hu[1]
[1] Shanghai Astronomical Observatory, Chinese Academy of Sciences, 80 Nandan Road, Shanghai 200030, People's Republic of China; htshang@shao.ac.cn, xhong@shao.ac.cn
[2] School of Physical Science and Technology, ShanghaiTech University, 100 Haike Road, Pudong, Shanghai 201210, People's Republic of China
[3] School of Astronomy and Space Science, University of Chinese Academy of Sciences, Beijing 100049, People's Republic of China
[4] College of Physical Science and Technology, Xiamen University, Xiamen 361005, People's Republic of China

## Abstract

We conducted a detailed analysis of the jet structure and dynamics of the source 0241+622 on milliarcsecond scales. We stacked images from multiple epochs to better recover the cross section of the jet. By analyzing the relationship between jet width and distance, we observed that the jet exhibits a parabolic shape from the core, spanning a region from 0.12 to 6.1 mas. This structure suggests the acceleration and collimation processes of the jet. Beyond 18 mas from the core, the jet adopts a conical shape, and the expansion speed of the jet becomes faster within the range from 4500 to 6500 mas. We obtained the core shift of this source using five pairs of data from the Very Long Baseline Array at 1.6 to 43 GHz. Based on previous studies, through proper motion analysis of the jet components, we estimated the angle between the jet and the line of sight to be approximately 65°.7, so 1 mas corresponds to 0.95 pc (deprojected distance). We then obtained the velocity field of the source within 3.14 mas from the central black hole and found that the jet exhibits accelerated motion within this range. At approximately 6.1 mas from the core, we observed that the jet width begins to decrease, which we identified as possibly corresponding to the Bondi radius of this source. The reduction in jet width may be related to changes in the external environmental pressure, particularly within the Bondi radius, indicating that the jet dynamics and collimation characteristics are strongly influenced by the surrounding medium conditions.

*Unified Astronomy Thesaurus concepts:* Active galactic nuclei (16); Relativistic jets (1390); Radio jets (1347)

## 1. Introduction

One of the most notable features of jets in active galactic nuclei (AGNs) is their length, which can extend up to 10 orders of magnitude greater than the size of the central black hole (M. Boettcher et al. 2012). However, the mechanisms by which AGN jets are collimated and accelerated have long been a matter of debate. Current theoretical studies and magnetohydrodynamic simulations suggest that AGN jets are accelerated by converting Poynting flux into kinetic energy (e.g., Z.-Y. Li et al. 1992; M. C. Begelman & Z.-Y. Li 1994; N. Vlahakis & A. Königl 2004), and their collimation is due to the pressure from the external medium (e.g., D. Eichler 1993; Y. Lyubarsky 2009). However, the observations are still insufficient, and the complexity introduced by relativistic average particle energies means that this explanation remains controversial (C. Casadio et al. 2021).

Very long baseline interferometry (VLBI) technology is a powerful tool for studying the formation, collimation, and acceleration of jets. This technique provides resolutions up to submilliarcseconds, allowing us to probe the innermost structures of relativistic jets in AGNs. The structure of the jet exhibits a complex evolutionary process. The region between $10^2 r_s$ and $10^5 r_s$ is dominated by magnetic fields, where the jet has a parabolic streamline and is accelerated; at $10^5 r_s$, an equilibrium between the energy densities of the magnetic field and the radiating particles is achieved, the jet transitions to a conical geometry, and acceleration stops (J. C. McKinney et al. 2012). Semianalytical models matching the warm jet to the ambient medium with a total electric current closed inside the jet also predict a transition from parabolic to conical shapes based on the Bondi pressure profile $P \propto r^{-2}$ (V. S. Beskin et al. 2017). Jet shape transitions from parabolic to conical have also been detected in 15 other nearby AGNs, such as NGC 315 (J. Park et al. 2021), NGC 6251 (C.-Y. Tseng et al. 2016), 1H 0323+342 (K. Hada et al. 2018), and NGC 1052 (S. Nakahara et al. 2020). Additionally, a transition from parabolic to cylindrical is detected in Cyg A (B. Boccardi et al. 2016; S. Nakahara et al. 2019). Y. Y. Kovalev et al. (2020) analyzed Very Long Baseline Array (VLBA) data from 367 AGNs at 15 and 1.4 GHz. They found that, in 29 nearby jets with redshifts $z < 0.07$, 10 jets transitioned from a parabolic shape to a conical shape. The full analysis sample primarily consists of distant AGNs with a redshift of $z \approx 1$. Therefore, they conclude that geometric transitions may be a common phenomenon in AGN jets and can be observed with sufficiently high-resolution observations.

0241+622 is a nearby Seyfert 1.2 galaxy ($z = 0.045$; M. P. Véron-Cetty & P. Véron 2006), whose central black hole has a mass of $\log_{10}\left(\frac{M_{bh}}{M_{sol}}\right) = 8.1^{+0.3}_{-0.3}$ (R. A. Phillipson et al. 2023). The source exhibits a distinct southeast-oriented jet (A. B. Pushkarev et al. 2017). Images at 43 GHz provided by X. P. Cheng et al. (2020) show the presence of a counterjet, whose flux density exceeds 7 times the rms noise ($\sigma_{rms}$), indicating that the detected structures are reliable and suggesting that the jet is oriented at a large angle relative to our line of sight. Based on the studies by M. L. Lister et al. (2019), the jet of 0241+622 can be modeled with two circular Gaussian components. The apparent speed of the inner component, at an average core distance of 0.79 mas, was 0.69$c$, while that of the outer component, at an average core distance of 2.89 mas, was 1.55$c$. Y. Y. Kovalev et al. (2020) used VLBA data at 15 and 1.4 GHz to search for jet shape transitions in 367 nearby AGNs, finding that only three sources exhibited purely parabolic jets, with 0241+622 being one of them. According to the results obtained by A. B. Pushkarev et al. (2017) using MOJAVE 15 GHz data, the jet extends up to about 5 mas from the core, and the shape of the jet begins to change at approximately 2.9 mas from the core. However, due to the limited observed jet length, its subsequent shape was not detected. Therefore, we attempt to use lower-frequency, specifically $X$-band data, to explore the changes in the jet shape of this source.

In this paper, we report the jet viewing angle constraints, velocity field, and collimation characteristics of the jet in 0241 +622 on parsec scales using multiepoch VLBI observational data. In Section 2, we describe the data used, as well as the data processing and analysis. In Section 3, we present the results, and in Section 4, we give a discussion of our findings.

Throughout the paper, we use a $\Lambda$CDM cosmology with $H_0 = 71$ km s$^{-1}$ Mpc$^{-1}$, $\Omega_m = 0.27$, and $\Omega_\Lambda = 0.73$ (E. Komatsu et al. 2011), and an angular size of 1 mas corresponds to a projected linear length of 0.87 pc at the redshift ($z = 0.045$) of 0241+622. With these conventions, 1 mas yr$^{-1}$ proper motion speed corresponds to 2.97$c$.

## 2. Data and Methods

### 2.1. Archival Data

To further analyze the properties of the jet in 0241+622, we searched for and obtained VLBA data suitable for studying jet collimation on the National Radio Astronomy Observatory (NRAO) archive[5] and Astrogeo VLBI FITS image database.[6] Since the 15 GHz data did not reveal any transitions in the jet shape, we plan to use lower-frequency data. We chose to use multiepoch data at 6.7 and 8 GHz for analysis, supplemented by a single-epoch 1.6 GHz VLBA image. Additionally, we downloaded publicly available stacked images of 0241+622 at 15 GHz for analysis, which are the same images presented by A. B. Pushkarev et al. (2017). To study the large-scale expansion profile of the source, we also analyzed the 1.5 GHz VLA calibrated data from the NRAO VLA Archive.[7]

In THE NRAO archive, some data are used for the delta zenith angle study and not for normal calibration or imaging. We excluded these data, as well as those without obvious jet extension structures. Based on this, we attempted to ensure that the data used were evenly distributed over time. We found over 40 epochs of 6.7 and 8 GHz VLBA observations in the NRAO archive. At the beginning, we excluded observations with only one or two scans of 0241+622, as they provided insufficient poor $uv$ coverage. Next, we processed the remaining observation data and, based on the resulting images, selected the data from six 6.7 GHz observation epochs (2012–2013) and five 8 GHz observation epochs (2006–2021) that best displayed the jet morphology. We applied the same criteria to filter the data in the Astrogeo database and ultimately obtained four epochs (2013–2017) of 8 GHz data. Overall, we acquired VLBA observation data from 16 epochs, covering 1.6, 6.7, and 8 GHz, spanning from 2006 to 2021. According to A. B. Pushkarev et al. (2017), the data we obtained from 16 epochs is sufficient for subsequent research. We have listed the detailed information about these data in Table 1. We also acquired four epochs of 1.5 GHz VLA observation data, and the detailed information is listed in Table 2.

In addition, to analyze the core shift of the source, we also obtained data from the NRAO archive, Astrogeo VLBI FITS image database, and MOJAVE database (M. L. Lister et al. 2018) for five frequency pairs ranging from 1.6 to 43 GHz. Detailed information about these data is listed in Table 3. We used data pairs with observation dates close to each other in order to minimize uncertainties caused by flux variations and plasma motion.

### 2.2. Data Analysis

The data in the Astrogeo VLBI FITS image database and the NRAO VLA Archive have already been fully self-calibrated. Therefore, we directly imaged these data using Difmap (M. C. Shepherd et al. 1994). The data from the NRAO archive were calibrated using the NRAO Astronomical Image Processing System (AIPS; E. W. Greisen 2003) following the standard procedures described in the AIPS cookbook[8] at first. Then we perform self-calibration (both phase and amplitude) and imaging loops using Difmap to obtain the final images. Natural weighting was used during the imaging.

First, we attempted to directly use the VLBA 8 GHz single-epoch images to analyze the jet shape, and the results are shown in Figure 1. We can observe that the jet shape is highly unstable; however, as the observation time increases, the shape of the jet becomes smoother. This may be because the increase in observation time effectively reduces the rms noise ($\sigma_{\rm rms}$) in the final image, allowing for better recovery of the complete cross section of the jet (A. B. Pushkarev et al. 2017; C. Casadio et al. 2021). We calculated the mean and standard deviation of the power-law indices describing the jet shape across the nine epochs to be 1.04 and 0.47, respectively. The coefficient of variation is approximately 45.3%, indicating a high level of variability, which exceeds the commonly used 30% threshold for data consistency in many scientific analyses (D. Montgomery & G. Runger 2010). We then stacked the images from the nine epochs to analyze the jet shape. The resulting power-law index that describes the jet shape is 0.61 ± 0.05, with a coefficient of variation of approximately 8.2%, which is significantly lower than the 45.3% obtained from the single-epoch measurements. This indicates that the measurement from the stacked image is substantially more stable compared to those from individual epochs. In this study, we ultimately used images obtained by stacking different epochs at the same frequency band for subsequent analysis. For more information on the stacked images, see the Appendix. It is important to note that the jet width used in this study is derived from the averaged brightness distribution obtained by stacking multiple observational data sets. Therefore, the measured width does not directly represent the instantaneous or spontaneous jet flow width, but rather reflects the time-averaged, historical overall structure of the jet flow region. We adopt this approach to achieve a higher signal-to-noise ratio and obtain a more stable measurement of the overall collimation profile, but we acknowledge that this

---

[5] https://data.nrao.edu/portal/#/
[6] https://astrogeo.org/vlbi_images/
[7] https://www.vla.nrao.edu/astro/nvas/
[8] http://www.aips.nrao.edu/cook.html

**Table 1**
Summary of 0241+622 Observations

| Code (1) | Obs. Date (2) | Array (3) | Frequency (GHz) (4) | Data Rate (Mbit s$^{-1}$) (5) | Time (s) (6) | $S_{tot}$ (Jy) (7) | $S_{peak}$ (Jy beam$^{-1}$) (8) | $\sigma_{rms}$ (mJy beam$^{-1}$) (9) | Synthesized Beam (mas, mas, degree) (10) |
|---|---|---|---|---|---|---|---|---|---|
| BK224C | 2019-09-22 | VLBA | 1.6 | 2048 | 2262 | 0.430 | 0.046 | 0.41 | 10.1, 8.93, −88.2 |
| BR149AA | 2012-07-13 | VLBA − MK, SC | 6.7 | 512 | 222 | 0.962 | 0.556 | 0.33 | 3.41, 2.22, −6.17 |
| BR149AB | 2012-08-07 | VLBA − HN, LA, KP, PT, MK | 6.7 | 512 | 225 | 0.975 | 0.587 | 0.75 | 3.15, 2.36, 59.4 |
| BR149S1 | 2012-12-08 | VLBA | 6.7 | 512 | 429 | 0.754 | 0.332 | 0.19 | 2.73, 1.77, 33.2 |
| BR149S2 | 2013-05-19 | VLBA − SC | 6.7 | 512 | 430 | 0.718 | 0.363 | 0.23 | 3.03, 2.21, 26 |
| BR149S3 | 2013-06-24 | VLBA − FD | 6.7 | 512 | 428 | 0.688 | 0.240 | 0.33 | 2.38, 1.4, 39.6 |
| BR149S4 | 2013-11-24 | VLBA − FD | 6.7 | 512 | 431 | 0.732 | 0.235 | 0.25 | 2.36, 1.2, 41.6 |
| BD105 | 2006-02-02 | VLBA | 8.4 | 128 | 5280 | 0.645 | 0.322 | 0.14 | 2.25, 1.25, −2.91 |
| SS001B | 2010-07-22 | VLBA − NL | 8.4 | 512 | 1646 | 1.809 | 1.000 | 0.65 | 2.63, 1.16, −29.6 |
| RV101 | 2013-09-11 | VLBA − BR | 8.6 | 128 | 256 | 0.860 | 0.555 | 0.64 | 2.62, 1.49, 82.8 |
| RV102 | 2013-12-11 | VLBA − BR, OV | 8.4 | 128 | 264 | 0.878 | 0.559 | 0.65 | 2.1, 1.42, 74.3 |
| BW114A | 2015-07-24 | VLBA | 8.4 | 512 | 3509 | 0.929 | 0.554 | 0.13 | 2.45, 1.34, −0.42 |
| RV122 | 2017-04-25 | VLBA − NL | 8.6 | 128 | 228 | 0.898 | 0.572 | 0.56 | 2.22, 1.31, 84.7 |
| UF001I | 2017-05-27 | VLBA | 8.7 | 1024 | 590 | 0.998 | 0.617 | 0.44 | 1.77, 1.46, −28.2 |
| RV124 | 2017-07-17 | VLBA | 8.4 | 128 | 256 | 0.935 | 0.490 | 0.72 | 1.68, 1.18, −31.4 |
| UH007F | 2021-02-26 | VLBA | 8.7 | 1024 | 174 | 0.417 | 0.264 | 0.41 | 3.08, 1.29, 40.1 |

**Note.** The columns are as follows: (1) project codes, (2) date of observation, (3) participating stations (additional or nonparticipating stations are indicated with a plus sign and a minus sign, respectively), (4) frequency, (5) aggregate data rate, (6) total observation time of the target source, (7) total flux density of the image, (8) peak Intensity, (9) rms noise, and (10) synthesized beam from naturally weighting.

Table 2
Log of Observations and Characteristics of the Archival VLA Data Sets of 0241+622

| Code | Obs. Date | Array | Configuration | Frequency (GHz) | Time (s) | $S_{tot}$ (Jy) | $S_{peak}$ (Jy beam$^{-1}$) | $\sigma_{rms}$ (mJy beam$^{-1}$) | Synthesized Beam (mas, mas, degree) |
|---|---|---|---|---|---|---|---|---|---|
| (1) | (2) | (3) | (4) | (5) | (6) | (7) | (8) | (9) | (10) |
| PERLEY | 1983-09-06 | VLA | A/A | 1.5 | 920 | 0.248 | 0.181 | 0.35 | 3650, 1360, −62.5 |
| AM0124 | 1986-02-25 | VLA | A/A | 1.5 | 580 | 0.249 | 0.179 | 0.36 | 2740, 1340, 77.8 |
|  | 1986-06-15 | VLA | B/A | 1.5 | 750 | 0.258 | 0.186 | 0.40 | 4660, 2260, 58.3 |
| TESTT | 1995-10-04 | VLA | B/A | 1.5 | 160 | 0.273 | 0.198 | 0.72 | 4780, 3240, 72.2 |

**Note.** The columns are as follows: (1) project codes, (2) date of observation, (3) participating stations, (4) array configuration, (5) frequency, (6) total observation time of the target source, (7) total flux density of the image, (8) peak Intensity, (9) rms noise, and (10) synthesized beam from naturally weighting.

method may smooth out transient features and temporal variability in jet width.

Before stacking the images, the same circular restoring beam was used for each epoch image at the same frequency (C. Casadio et al. 2021), where the size of the restoring beam is the median beam size of all epochs (A. B. Pushkarev et al. 2017). Alignment can be achieved using the core components obtained from model fitting, assuming they are stationary. Although this may not coincide with the true jet base, this feature is the most compact and describes an area where morphological changes are negligible (B. Boccardi et al. 2016). Alternatively, alignment can be directly based on the position of the peak intensity in each image, and after aligning the total intensity maps of individual epochs, a simple average is performed on the image plane (C. Casadio et al. 2021). In this study, we used restoring beam sizes of 2.36, 1.59, and 2990 mas for the stacked images at 6.7, 8, and VLA 1.5 GHz, respectively, and aligned these images using the position of the peak intensity in each image. The $\sigma_{rms}$ of the stacked images were calculated using the method described in A. B. Pushkarev et al. (2017).

We used the stacked 6.7 and 8 GHz VLBA images, the single-epoch 1.6 GHz VLBA images, and the stacked 1.5 GHz VLA images to analyze the collimation profile of the 0241 +622 jet. We first performed model fitting on these images. Since the jet direction of the source showed little change over the time range of the data we used, we took the average of the position angles (PA = 122°.6) obtained from the model fitting of the jet components in each epoch—the result is displayed in Figure 2—and made vertical cuts along this value (S. Nakahara et al. 2020). For the 6.7 and 8 GHz data, we employed our custom Python script to slice the jet from the core downstream in 0.2 mas steps (H. Ro et al. 2023), obtaining the one-dimensional brightness distribution of the jet. We used the same approach for the 1.6 and VLA 1.5 GHz data. However, because its scale is larger, we followed the method described by K. Yi et al. (2024), starting the slicing from a point 1.5 times the restoring beam size away from the core and using a step size equal to one-third of the restoring beam size. Next, we used Gaussian distribution fitting for the brightness distribution of the jet at each slice. For fitting with a single Gaussian component, we used the following formula to determine the jet width (W; C. Casadio et al. 2021):

$$W = \sqrt{FWHM^2 - b^2}, \quad (1)$$

where FWHM is the full width at half-maximum from the single Gaussian fitting and $b$ is the size of the restoring beam perpendicular to the direction of the jet. For dual Gaussian components, the jet width at each slice is the distance between the edges of the two (deconvolved) FWHMs (B. Boccardi et al. 2016). Moreover, determining whether to use dual Gaussian components for fitting at a certain point in the jet involves considering positional errors, which can be assumed to be one-fifth of the FWHM of each Gaussian component (B. Boccardi et al. 2016). We found that the jet of 0241+622 is best fit by a single Gaussian component across all ranges.

Referring to S. Nakahara et al. (2020), we first obtain the error in the FWHM from the Gaussian fitting and then calculate the error in the jet width through error propagation. The sources of the fitting error for the Gaussian function FWHM include image noise, deconvolution errors from the imaging process, and deviations between the slice profiles and the Gaussian model (S. Nakahara et al. 2020). The criteria for successful Gaussian fitting are as follows: (1) the brightness of the jet exceeds 9 times the image noise, allowing for reliable measurements only at bright locations within sparse intensity distributions (H. Okino et al. 2022); (2) FWHM > $b$, ensuring that the jet radiation in the perpendicular direction is well resolved (H. Okino et al. 2022); and (3) the error of the calculated jet width does not exceed 100% of the actual width value (S. Nakahara et al. 2018). After fitting, connecting the Gaussian peaks at each slice position results in the jet ridge line. In previous studies, there were various definitions of the AGN jet ridge line, such as S. Britzen et al. (2017), M. Perucho et al. (2012), C. M. Fromm et al. (2013a), M. H. Cohen et al. (2015), and A. B. Pushkarev et al. (2017). Our research does not focus on analyzing the changes in the jet ridge line; the jet ridge line is naturally obtained during the process of slicing the jet and performing Gaussian fitting, so we do not discuss the issue of jet ridge lines in depth.

## 3. Results

### 3.1. Source Morphology

We present the naturally weighted VLBA images of 0241 +622 at 1.6, 6.7, and 8 GHz in Figures 3, 4, and 5, respectively. We also present the naturally weighted VLA image of the source at 1.5 GHz in Figure 6. From the figures, we can see that the source has a compact core region and a southeast-directed jet structure. The jet direction and morphology are relatively stable, showing no significant spiral motion or precession of the jet nozzle. We present the VLBA 1.6 GHz single-epoch image, the composite stacked VLBA images at 6.7 and 8 GHz, and the VLA 1.5 GHz composite stacked image along with the corresponding jet ridge lines in Figures 7 and 8. The rms noise ($\sigma_{rms}$) of the stacked images is 0.11, 0.12, and 0.33 mJy beam$^{-1}$, respectively, showing a

Table 3
Summary of Observations Used to Obtain the Core Shift of 0241+622

| Code | Obs. Date | Array | Frequency (GHz) | Data Rate (Mbit s$^{-1}$) | Time (s) | $S_{\rm tot}$ (Jy) | $S_{\rm peak}$ (Jy beam$^{-1}$) | $\sigma_{\rm rms}$ (mJy beam$^{-1}$) | Synthesized Beam (mas, mas, degree) |
|---|---|---|---|---|---|---|---|---|---|
| (1) | (2) | (3) | (4) | (5) | (6) | (7) | (8) | (9) | (10) |
| BG246AB | 2017-07-22 | VLBA | 1.6 | 2048 | 202 | 0.489 | 0.076 | 1.02 | 19.3, 7.63, 89.3 |
| BR149S4 | 2013-11-24 | VLBA − FD | 6.7 | 512 | 431 | 0.732 | 0.235 | 0.25 | 2.36, 1.2, 41.6 |
| RV124 | 2017-07-17 | VLBA | 8.4 | 128 | 256 | 0.935 | 0.490 | 0.72 | 1.68, 1.18, −31.4 |
| RV102 | 2013-12-11 | VLBA − BR, OV | 8.4 | 128 | 264 | 0.878 | 0.559 | 0.65 | 2.1, 1.42, 74.3 |
| BD105 | 2006-02-02 | VLBA | 8.4 | 128 | 5280 | 0.645 | 0.322 | 0.14 | 2.25, 1.25, −2.91 |
| BL137C | 2006-04-05 | VLBA | 15.4 | 256 | 501 | 0.596 | 0.416 | 0.77 | 0.79, 0.69, 38.3 |
| BL178AX | 2012-12-23 | VLBA | 15.4 | 512 | 2184 | 0.998 | 0.686 | 0.34 | 0.94, 0.69, 27.2 |
| BB313AJ | 2012-12-20 | VLBA + GB | 22.1 | 512 | 86 | 0.792 | 0.633 | 2.42 | 1.26, 0.34, 54.5 |
| BR210CC | 2015-11-23 | VLBA − SC | 22.1 | 512 | 63 | 1.378 | 1.250 | 1.88 | 1.44, 0.63, 84.3 |
| BA111J | 2015-11-30 | VLBA | 43.1 | 2048 | 1052 | 0.849 | 0.735 | 0.41 | 0.51, 0.32, −36.5 |

**Note.** The columns are as follows: (1) project codes, (2) date of observation, (3) participating stations (additional or nonparticipating stations are indicated with a plus sign and a minus sign, respectively), (4) frequency, (5) aggregate data rate, (6) total observation time of the target source, (7) total flux density of the image, (8) peak Intensity, (9) rms noise, and (10) synthesized beam from naturally weighting.

significant reduction in noise compared to the images before stacking.

### 3.2. Core Shift

Core shift effect refers to the phenomenon where the position of the core in the jets of AGNs changes due to the frequency dependence of synchrotron self-absorption (R. D. Blandford & A. Königl 1979; G. B. Rybicki & A. P. Lightman 1979). This effect has been observed in many blazars and radio galaxies (J. Park et al. 2021). By accurately measuring it at multiple frequencies, the position of the central engine can be inferred, which is of great importance for studying the structure of jets (K. Hada et al. 2011; J. Park et al. 2021).

To obtain the core offset between two frequencies, we used a two-dimensional cross-correlation method for the optically thin emission regions in jet images, based on the approach described by S. M. Croke & D. C. Gabuzda (2008). This method is built on a well-validated assumption that the emitting regions from the extended jet are optically thin features whose positions do not change at different frequencies. By aligning the corresponding images of these optically thin emission regions, the core offset between different frequencies can be determined (H. Okino et al. 2022). First, we set the pixel size of the two images to one-twentieth of the restoring beam size at the lower frequency, ensuring it is much smaller than the restoring beam size of the low-frequency image. Next, we convolved both images with the restoring beam of the lower frequency. Subsequently, we masked the optically thick core regions in both images and calculated the cross-correlation coefficient to determine the core offset for each frequency pair. We assume that the uncertainty of the core shift is 1/20 of the restoring beam (C. M. Fromm et al. 2013b; A. M. Kutkin et al. 2014). The results are summarized in Table 4.

After obtaining the core shifts for each frequency pair, we used the 43 GHz core position as a reference and performed a least-squares fit to the core positions at each frequency relative to the 43 GHz core position using the following equation (J. Park et al. 2021):

$$\Delta r_{\rm core} = a\nu^b + c. \quad (2)$$

Here, $\Delta r_{\rm core}$ represents the distance of the core position at each frequency relative to the 43 GHz core position, and $\nu$ represents the frequency. Figure 9 shows the fitting results, where $a = 8.10 \pm 1.27$, $b = -1.12 \pm 0.09$, and $c = -0.12 \pm 0.03$. By extrapolating the frequency to infinity, we determined that the jet base is located approximately 0.12 mas upstream from the 43 GHz core.

### 3.3. Jet Viewing Angle Constraints and Velocity Field

M. L. Lister et al. (2019) used MOJAVE 15 GHz data from 2003 to 2012 to perform a model fit on 0241+622, discovering that the jet of the source consists of two components with apparent velocities ($\beta_{\rm app}$) of $1.55 \pm 0.09c$ and $0.69 \pm 0.03c$, respectively. The average distances of these components from the core are 2.89 and 0.79 mas, respectively, indicating that the jet components are in accelerating motion within the observed range. Based on the results obtained above, the intrinsic Lorentz factors ($\Gamma$) and the angle between jet and line of sight of the observer ($\theta$) were calculated using the method described in X. Li et al. (2018) for these two jet components. According to Equations (1) and (2) in their paper, we calculated the maximum apparent velocity ($\beta_{\rm app} = 1.55c$) of the source jet to obtain $\Gamma \geqslant 1.84$ and $\theta \leqslant 65°.7$.

Since the 43 GHz images of the source show the presence of a counterjet (X. P. Cheng et al. 2020), we can use the following formula to determine the jet velocity, according to J. Park et al. (2021):

$$R \equiv \frac{I_{\rm jet}}{I_{\rm cjet}} = \left(\frac{1 + \beta \cos\theta}{1 - \beta \cos\theta}\right)^{2-\alpha}, \quad (3)$$

where $I_{\rm jet}$ and $I_{\rm cjet}$ are the intensities of the jet and counterjet at the same distance, $\beta$ is the intrinsic velocity of the jet, and $\alpha$ represents the spectral index of synchrotron radiation. For 0241+622, the radio spectrum is quite flat (T. Hovatta et al. 2014), indicating $\alpha \sim 0$.

The intensities of the jet and counterjet were obtained by fitting a single Gaussian function to the transverse intensity distribution of the jet at each distance. To avoid potential contamination from the bright core emission, we used data at positions greater than one synthesized beam major axis away from the core for the analysis, with a step size of half the major axis of the restoring beam (J. Park et al. 2021). After considering the core shift effect described in Section 3.2, we started slicing from a distance of 0.51 mas from the 43 GHz core. Due to the short length of the counterjet at 43 GHz for this source, we only obtained the brightness ratio of the jet and counterjet at a distance of 0.51 mas from the 43 GHz core, which is $R = 1.94$. Assuming $\theta = 65°.7$, we find $\beta = 0.40c$. Then, using the following formula, we obtained the intrinsic velocities corresponding to the apparent velocities in M. L. Lister et al. (2019):

$$\beta_{\rm app} = \frac{\beta \sin\theta}{1 - \beta \cos\theta}, \qquad (4)$$

which are $0.58 \pm 0.02c$ and $1.00 \pm 0.14c$, respectively, from near to far distances from the core. We present the relationship between the distance from the central black hole and the intrinsic velocity of the jet in Figure 10. We can observe that the jet may accelerate within the range displayed in the figure. This is consistent with our fitting results for the jet profile of this source in the next section, as the acceleration region of the jet is often referred to as the acceleration and collimation zone (e.g., A. P. Marscher et al. 2008). In this zone, the jet typically has a parabolic shape, and it may also be a more collimated cylindrical shape (C. Casadio et al. 2021).

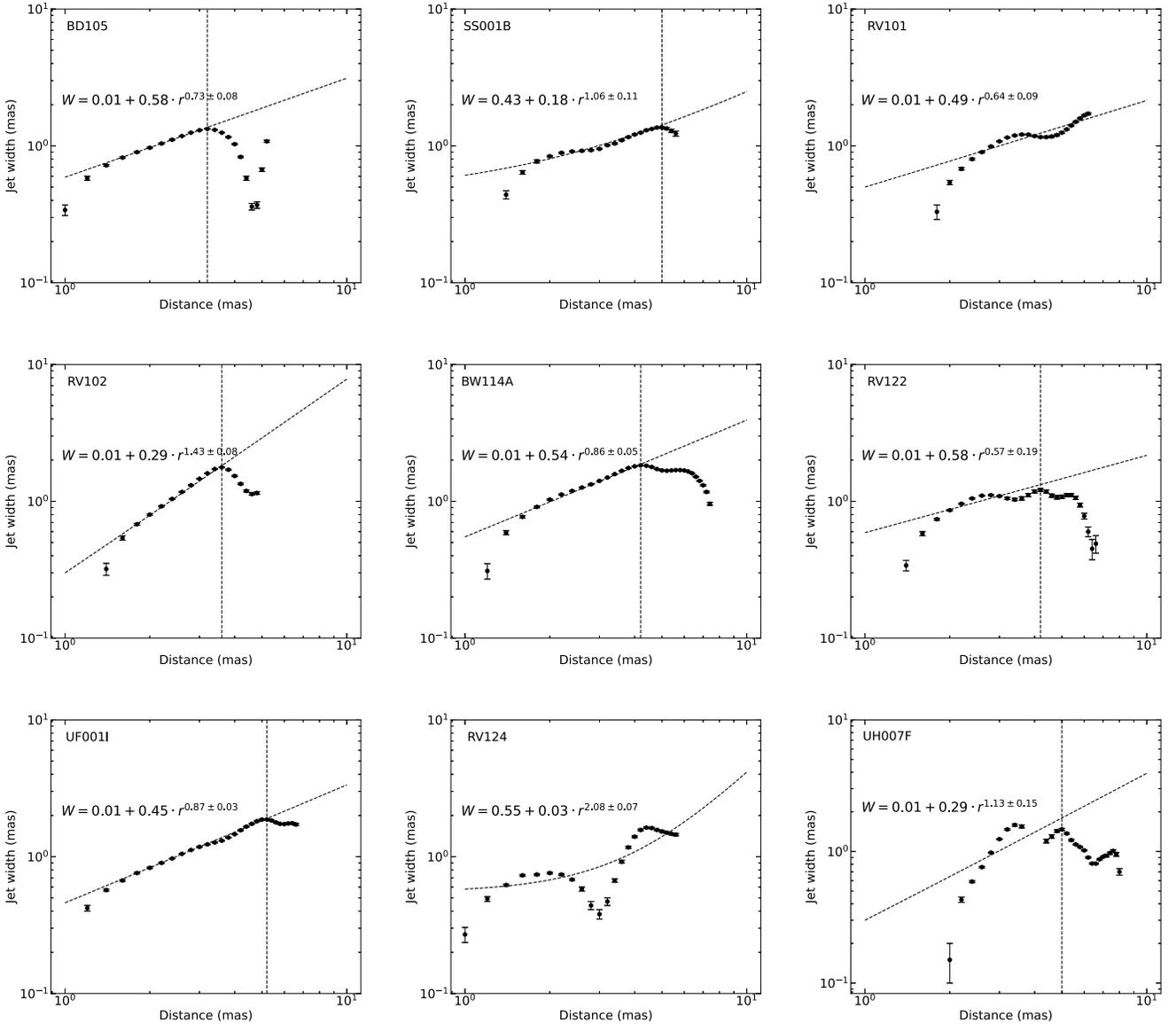

**Figure 1.** The jet width vs. radial distance for the nine epochs at VLBA 8 GHz. The dashed diagonal line represents the best-fit curve, and the vertical dashed line indicates the position where the jet begins to narrow.

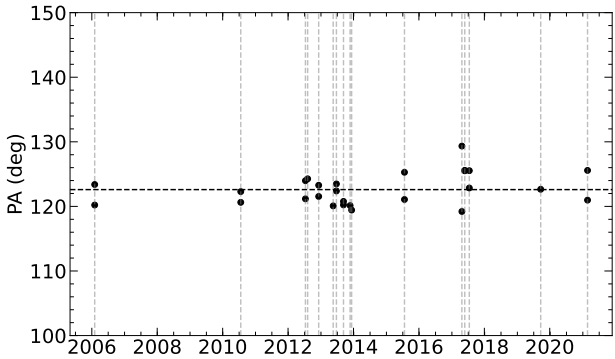

**Figure 2.** The results of model fitting for the jet data across 15 epochs, with the horizontal axis representing time and the vertical axis showing the position angle of the jet components obtained from the fitting. The black dashed line indicates the average position angle (PA = 122°.6) of all jet components.

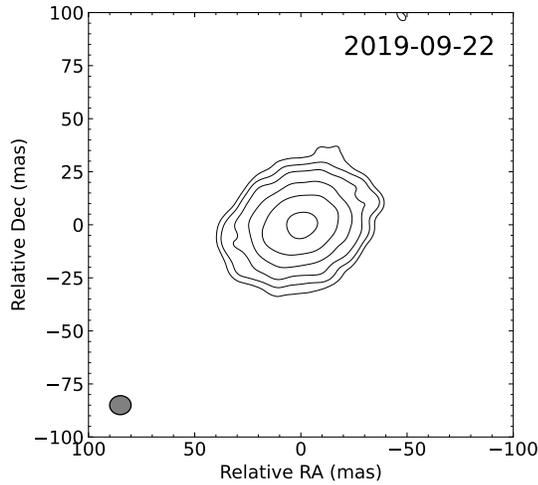

**Figure 3.** The naturally weighted images of 0241+622 were acquired during the 1.6 GHz VLBA observations. Contours start at 3 times the rms noise ($\sigma_{rms}$) level and increase in steps of 2. The gray-filled ellipses in the lower left indicate the synthesized beam for each image. Detailed information about this image can be found in Table 1.

D. C. Homan et al. (2021) calculated the angle between the jet of the source and the line of sight to be 49°.8 using the median estimated intrinsic brightness temperature, which is consistent with our result. The data used also came from M. L. Lister et al. (2019). The median estimated intrinsic brightness temperatures obtained using different methods vary (e.g., T. Hovatta et al. 2009; I. Liodakis et al. 2017; D. C. Homan et al. 2021), and the resulting angles between the jet and the line of sight are also affected. For 0241+622, which is relatively close to us ($z = 0.045$) and shows a counterjet in its 43 GHz images (X. P. Cheng et al. 2020), this indicates a larger angle between the jet and our line of sight. In such cases, slight variations in the angle between the jet and the line of sight do not affect our subsequent analysis. In the following analysis, we take the angle between the jet of the source and the line of sight to be 65°.7. Therefore, 1 mas corresponds to 0.95 pc (deprojected distance).

### 3.4. Jet Collimation Profile

We utilized the 8, 15, 22, and 43 GHz data sets to determine the core shift of the source. By fitting a Gaussian model to the innermost bright region in each image, we derived core sizes of $0.29 \pm 0.03$, $0.15 \pm 0.01$, $0.12 \pm 0.02$, and $0.07 \pm 0.01$ mas at 8, 15, 22, and 43 GHz, respectively (M. Nakamura & K. Asada 2013). These measurements were further analyzed in conjunction with the jet cross-sectional profiles obtained from the slicing analysis. We show the best-fit parameters for the jet shape and their relative uncertainties in Figure 11. We used nonlinear least-squares fitting to model the relationship between the jet width and distance, considering the core shift and aligning the data from all frequencies with respect to the central black hole. From the results at 6.7 and 8 GHz, we observe that the jet width of this source starts to decrease at 6.68 and 5.53 mas, respectively. According to the results at 1.6 GHz, we observe that the jet begins to expand again as the distance increases. We have observed that the phenomenon of the jet width decreasing for this source has also been observed in 3C 84 (G. Giovannini et al. 2018) and 1H 0323+342 (K. Hada et al. 2018), where the jet width decreases in the recollimation zone, and then the expansion rate increases, forming a conical shape. For 0241+622, it is not possible to describe the overall variation of the jet width using a single power-law relation between $W - r$. Therefore, we take the average position where the jet width begins to decrease as $r = 6.1$ mas. For the range from 0.12 to 6.1 mas, we use $W = a + c \cdot r^k$ for fitting, where $a$ gives the jet width at the location of the core. For the jet range beyond 18 mas, we use $W = c \cdot r^k$ for fitting. Additionally, for the VLA 1.5 GHz results, we also fit them separately using $W = c \cdot r^k$ (C. Casadio et al. 2021).

In the range from 0.12 to 6.1 mas, our fitting result is $k = 0.72 \pm 0.01$. Although the relationship between $k$ and the jet shape is not consistently defined in different articles, such as K. Yi et al. (2024) and B. Boccardi et al. (2021), our fitting results clearly indicate that the jet of this source exhibits a parabolic shape within the range before 6.1 mas. This result is consistent with the findings of Y. Y. Kovalev et al. (2020), although they only fitted the jet within the range of 0.5–2.9 mas. A. B. Pushkarev et al. (2017) used MOJAVE 15 GHz data from 2003 to 2012 to fit the jet shape of the source from 0.5 to 5 mas, obtaining $k = 1.09 \pm 0.08$. However, their fitting was relatively rough, and at this observational frequency, the jet radiation becomes quite weak around 3 mas. Not excluding this region during the fitting likely resulted in a faster apparent expansion of the jet, thus giving it a conical shape. For the range beyond 18 mas, our fitting result is $k = 1.04 \pm 0.01$. This indicates that the jet takes a conical shape in this range. Additionally, the fitting result for the VLA 1.5 GHz data is $k = 2.97 \pm 0.13$, which indicates that the expansion speed of the jet becomes faster at this scale, with a range approximately from 4500 to 6500 mas ($3.8 \times 10^8 r_s$ to $5.5 \times 10^8 r_s$). This range is close to the region where the jet becomes diffuse ($10^9 r_s$; B. Boccardi et al. 2017).

### 4. Discussion

In this study, we used multiepoch stacked data supplemented with lower-frequency single-epoch data to investigate the jet morphology of 0241+622. We found that within the range from 0.12 to 6.1 mas from the core, the jet width increases gradually and follows a parabolic shape. Beyond 6.1 mas, the width of the jet gradually decreases. The jet then begins to reexpand beyond 18 mas, adopting a conical shape. Theoretical models and simulations of magnetically driven jets, transitioning from Poynting flux to kinetic flows, indicate that the properties of the external medium represent a fundamental parameter (S. S. Komissarov et al. 2007; A. Tchekhovskoy et al. 2008; Y. Lyubarsky 2009). Specifically, the external pressure gradient along the direction of the jet largely determines the jet shape and acceleration, both of which are closely related.

The jet of 0241+622 expands in a parabolic shape within 6.1 mas of the core, undergoing acceleration and collimation in this region. The morphology of the flow within this area is determined by the magnetic field strength and the external pressure at the flow base. The source of the external pressure profile is likely the environmental medium confined by the gravitational field of the black hole, which traps the jet (K. Yi et al. 2024). Specifically, when the $\frac{6\pi\rho_0}{B_0^2}$ ratio exceeds 0.25, the flow approximates a steady state, forming a perfect parabolic shape, where the exponent of the relationship between the jet radius and the distance from the core is 0.5 (Y. Lyubarsky 2009). When the ratio is smaller, the flow approximates a nonequilibrium state (Y. Lyubarsky 2009). Simultaneous acceleration and collimation can be explained as "differential collimation" (also known as the magnetic nozzle effect; e.g., M. Camenzind 1987; Z.-Y. Li et al. 1992; M. C. Begelman & Z.-Y. Li 1994). This mechanism enables the axial magnetic field lines to focus inward and diverge outward simultaneously. This field structure can act as a nozzle, effectively converting Poynting flux into kinetic energy (S. S. Komissarov 2011; M. Nakamura & K. Asada 2013). Energy conversion facilitates efficient jet acceleration, leading to linear acceleration: the bulk Lorentz factor $\Gamma$ increases linearly with the jet radius ($\Gamma \propto R$; V. S. Beskin & E. E. Nokhrina 2006; A. Tchekhovskoy et al. 2008; Y. Lyubarsky 2009).

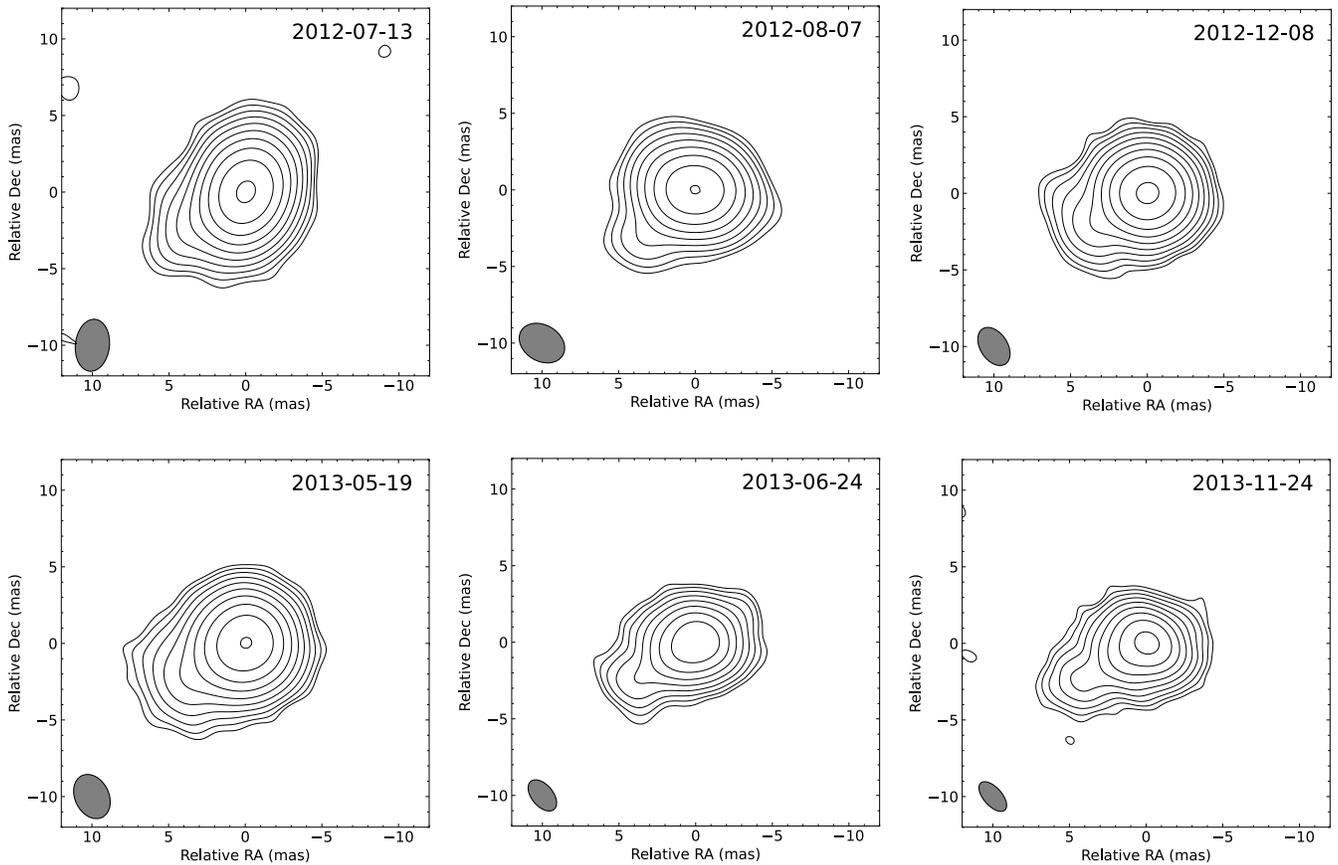

**Figure 4.** The naturally weighted images of 0241+622 were acquired during the 6.7 GHz VLBA observations from 2012 to 2013. Contours start at 3 times the rms noise ($\sigma_{\rm rms}$) level and increase in steps of 2. The gray-filled ellipses in the lower left indicate the synthesized beam for each image. Detailed information about these images can be found in Table 1.

For 0241+622, beyond a distance of 6.1 mas from the core, the jet width begins to decrease. This may be due to a change in external pressure at the Bondi radius, altering the jet expansion rate. Within the Bondi radius, the external medium pressure decreases with increasing distance from the black hole, while beyond the Bondi radius, the pressure profile becomes flatter, consistent with the general pressure profile of the host galaxy. The jet inertia causes it to overexpand, after which its pressure becomes insufficient relative to the external medium, resulting in recollimation shocks (C. Casadio et al. 2021). In purely hydrodynamic scenarios, recollimation naturally occurs when the expanding fluid is underpressured compared to the surrounding medium (R. A. Daly & A. P. Marscher 1988). In magnetohydrodynamic flows, collimation features also emerge when the jet is no longer strongly magnetized and becomes causally disconnected from the central engine (D. L. Meier 2012). Subsequently, the jet will transition into other shapes.

Regarding the scenario of jet collimation breaks, Y. Y. Kovalev et al. (2020) have pointed out that under a single environmental pressure distribution, such as $p \propto z^{-2}$, jet collimation breaks occur in regions where magnetic energy and particle energy are equal. They have developed a model in which the location of the jet collimation break primarily depends on the magnetization parameter at the jet base ($\sigma_M$), as well as the mass and spin of the central black hole (E. E. Nokhrina et al. 2020). However, it is currently challenging to strictly constrain the values of the magnetization parameter ($\sigma_M$; H. Okino et al. 2022).

The Bondi radius, which demarcates the extent of a black hole's gravitational influence, typically extends to $10^5 \sim 10^6 r_s$ (K. Asada & M. Nakamura 2012; G. Boccardi et al. 2020). This may be the cause of the significant changes in the surrounding pressure environment where the jet propagates. Based on the jet and line of sight angle of 0241+622 obtained in Section 3.3, we can calculate that 1 mas corresponds to 0.95 pc (deprojected distance). Based on the central black hole mass of the source provided by R. A. Phillipson et al. (2023; $\log_{10}\left(\frac{M_{\rm bh}}{M_{\rm sol}}\right) = 8.1^{+0.3}_{-0.3}$), we can use $r_s = \frac{2GM}{c^2}$ to calculate that its

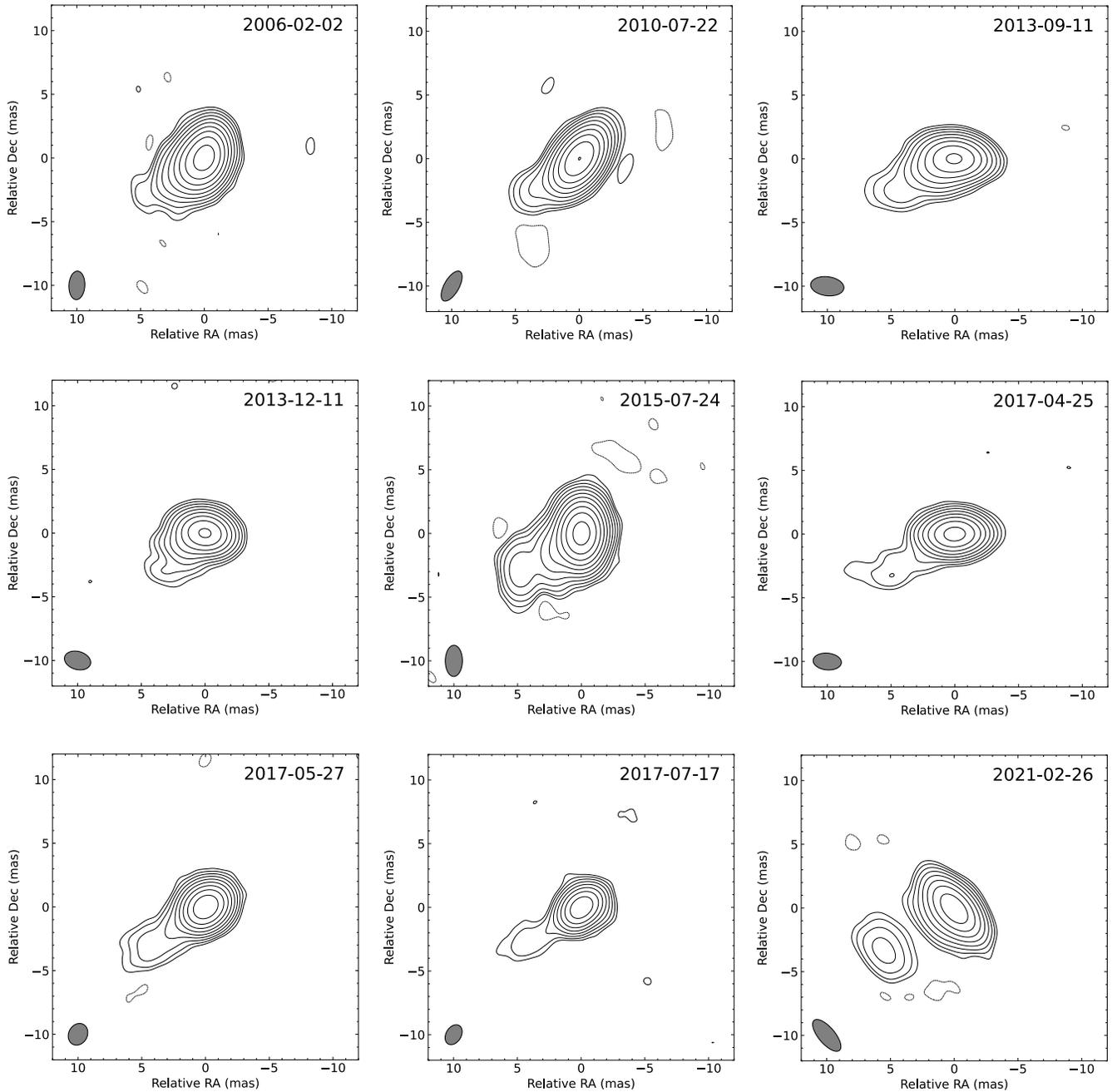

**Figure 5.** The naturally weighted images of 0241+622 were acquired during the 8 GHz VLBA observations from 2006 to 2021. Contours start at 3 times the rms noise ($\sigma_{\rm rms}$) level and increase in steps of 2. The gray-filled ellipses in the lower left indicate the synthesized beam for each image. Detailed information about these images can be found in Table 1.

Schwarzschild radius ($r_s$) is approximately $1.2 \times 10^{-5}$ pc. Thus, $10^5 \sim 10^6 r_s$ corresponds to $1.26 \sim 12.6$ mas. This value is consistent with our observational results, which show that the jet width begins to decrease at 6.1 mas.

Based on lower-frequency (1.5 GHz) images, we observed the morphology of the source on a larger scale. According to the results of J. B. Hutchings et al. (1982), at VLA 1.5 GHz there is a bright jet component approximately 5″ to the southeast of the core, with very weak jet components in between. This may be due to jet acceleration occurring within the Bondi radius (M. Nakamura & K. Asada 2013; K. Asada et al. 2014; F. Mertens et al. 2016; K. Hada et al. 2017; R. C. Walker et al. 2018; J. Park et al. 2019), transitioning to slower deceleration beyond the Bondi radius (J. A. Biretta et al. 1995; J. A. Biretta et al. 1999; E. T. Meyer et al. 2013). Beyond 6.1 mas, the jet slows down due to interactions with the surrounding medium. At this point, a change in the external pressure profile triggers the end of jet collimation, where the external pressure no longer provides sufficient support to confine the jet (K. Asada & M. Nakamura 2012; C.-Y. Tseng et al. 2016). Until about 5″ from the core, the external medium may become denser, and the interaction between the jet and this medium generates shock waves, thereby locally dissipating kinetic energy and producing bright, compact emission points, which correspond to the observed bright jet features (M. Boettcher et al. 2012). The environment around the cores of AGNs is extremely complex, and the nature of the medium that confines the jets in this region is still far from understood (B. Boccardi et al. 2021). Closer to the black hole, it may consist of the accretion disk itself or dense gas clouds surrounding it, while further away, it could involve contributions from the shocked cocoon around the jet or the interstellar medium. Additionally, disk winds from the accretion disk may effectively confine the internal relativistic jets (S. Bogovalov & K. Tsinganos 2005; N. Globus & A. Levinson 2016). Furthermore, different energy outputs could result in varied feedback on the environment (T. M. Heckman & P. N. Best 2014).

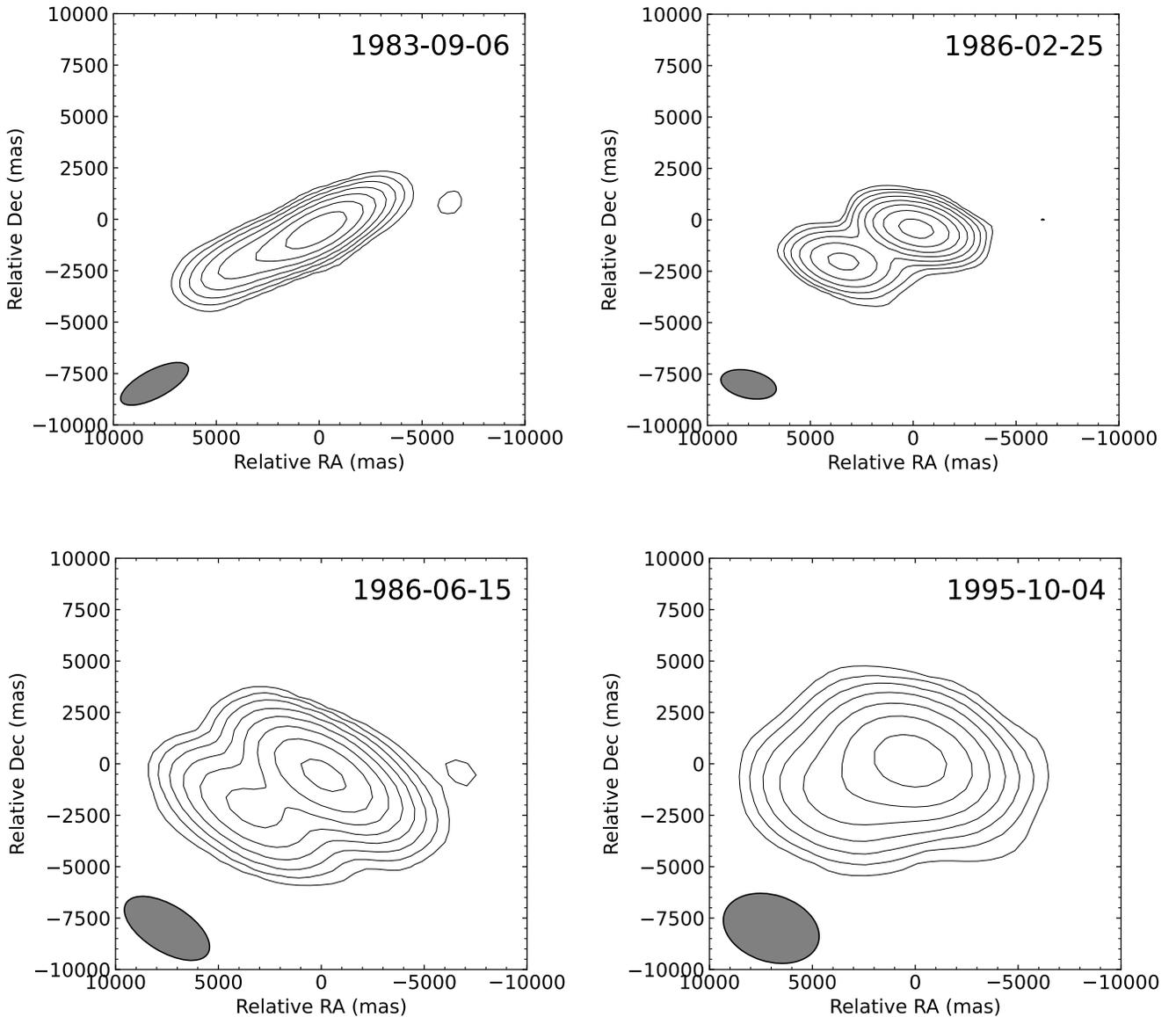

**Figure 6.** The naturally weighted images of 0241+622 were acquired during the 1.5 GHz VLA observations. Contours start at 3 times the rms noise ($\sigma_{\rm rms}$) level and increase in steps of 2. The gray-filled ellipses in the lower left indicate the synthesized beam for each image. Detailed information about these images can be found in Table 2.

Many radio-loud AGN jets, such as 3C 84 (G. Giovannini et al. 2018) and 3C 264 (B. Boccardi et al. 2019), are collimated within $\lesssim 10^4 R_g$ and transition downstream to a free-expansion phase, such as NGC 6251 (C.-Y. Tseng et al. 2016), 3C 273 (K. Akiyama et al. 2018; H. Okino et al. 2022), NGC 4261 (S. Nakahara et al. 2018), 1H 0323+342 (K. Hada et al. 2018), NGC 1052 (S. Nakahara et al. 2020), TXS 2013 +370 (E. Traianou et al. 2020), NGC 315 (J. Park et al. 2021), 3C 264 (B. Boccardi et al. 2019), and other sources (Y. Y. Kovalev et al. 2020, B. Boccardi et al. 2021; P. R. Burd et al. 2022). H. Okino et al. (2022) demonstrated the relationship between black hole mass and the deprojected distance of jet collimation breaks in 3C 273 and other AGN sources. They found that the break positions are broadly distributed between $10^4 \sim 10^8\, r_s$. Despite this, apart from a few sources, jet acceleration within the collimation zones has largely been unexplored. This is partly due to a lack of monitoring observations. In fact, reliable analysis of the jet velocity field requires multiepoch VLBI observations to be conducted at a high cadence (K. Yi et al. 2024). For sources where the counterjet has been observed, the jet-to-counterjet brightness ratio can also be used as an alternative method to multiepoch VLBI observations (J. Park et al. 2021).

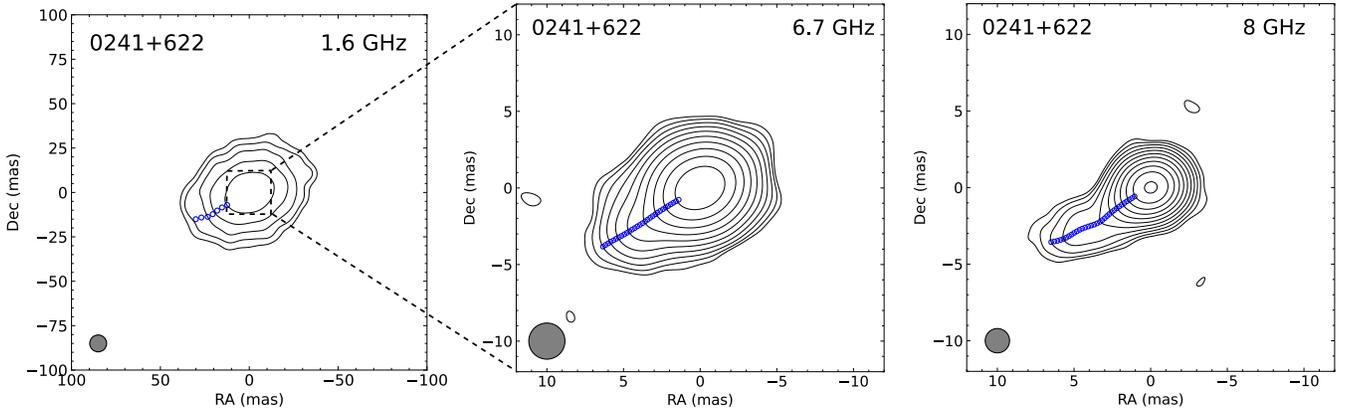

**Figure 7.** The single-epoch VLBA image at 1.6 GHz and the stacked VLBA images of 0241+622 at 6.7 and 8 GHz. The contours start from the lowest levels ($3\sigma_{rms}$) of 1.23, 0.33, and 0.36 mJy beam$^{-1}$, increasing by a factor of 2. The blue circles indicate the peaks of the fitted Gaussians.

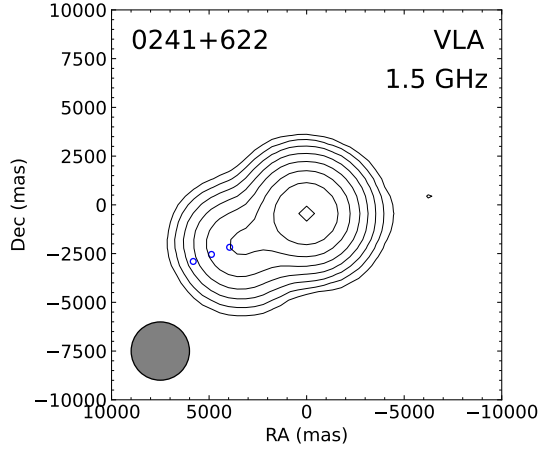

**Figure 8.** The stacked VLA images of 0241+622 at 1.5 GHz. The contours start from the lowest levels ($3\sigma_{rms}$) of 0.99 mJy beam$^{-1}$, increasing by a factor of 2. The blue circles indicate the peaks of the fitted Gaussians.

**Table 4**
Core Shifts between Each Pair of Frequencies

| $\nu_1$ (GHz) (1) | $\nu_2$ (GHz) (2) | $\Delta r$ (mas) (3) | Angle (deg) (4) |
|---|---|---|---|
| 22.1 | 15.4 | 0.15 ± 0.040 | 128.80 |
| 43.1 | 22.1 | 0.085 ± 0.045 | 110.54 |
| 15.4 | 8.4 | 0.39 ± 0.068 | 118.30 |
| 8.4 | 6.7 | 0.65 ± 0.084 | 144.83 |
| 1.6 | 8.4 | 3.87 ± 0.61 | 163.81 |

**Note.** The columns are as follows: (1) and (2) observing frequencies, (3) core shift, and (4) direction of the core shift.

## 5. Summary

In this paper, we present the kinematics and collimation characteristics of the jet in 0241+622 on parsec scales using multiepoch VLBI observational data. Our findings are summarized as follows:

1. By stacking multiple epochs of observational data, we obtained high dynamic range composite images. Through nonlinear least-squares fitting, we determined the relationship between jet width and distance, finding that the jet of 0241+622 exhibits a parabolic shape in the range of 0.12–6.1 mas, a conical shape beyond 18 mas, and at further distances (4500–6500 mas), the jet expansion speed increases. The corresponding fitting results give $k$ values of $0.72 \pm 0.01$, $1.04 \pm 0.01$, and $2.97 \pm 0.13$, respectively.

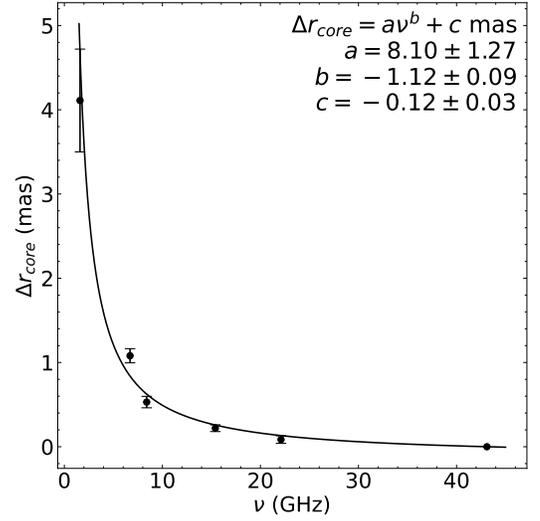

**Figure 9.** The variation of the core position of 0241+622 with frequency, using the 43 GHz core position as the reference point. The black solid line represents the best-fit power-law function, with the best-fit parameters noted in the upper-right corner.

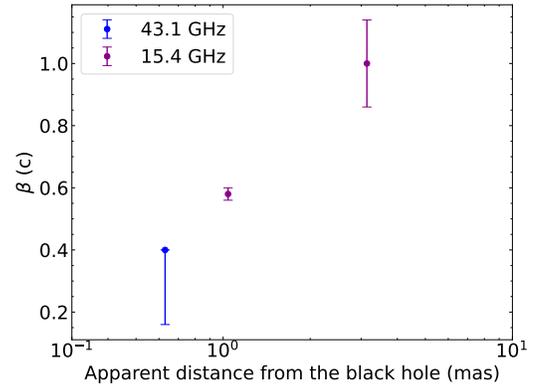

**Figure 10.** The intrinsic velocity of the jet as a function of apparent distance from the black hole.

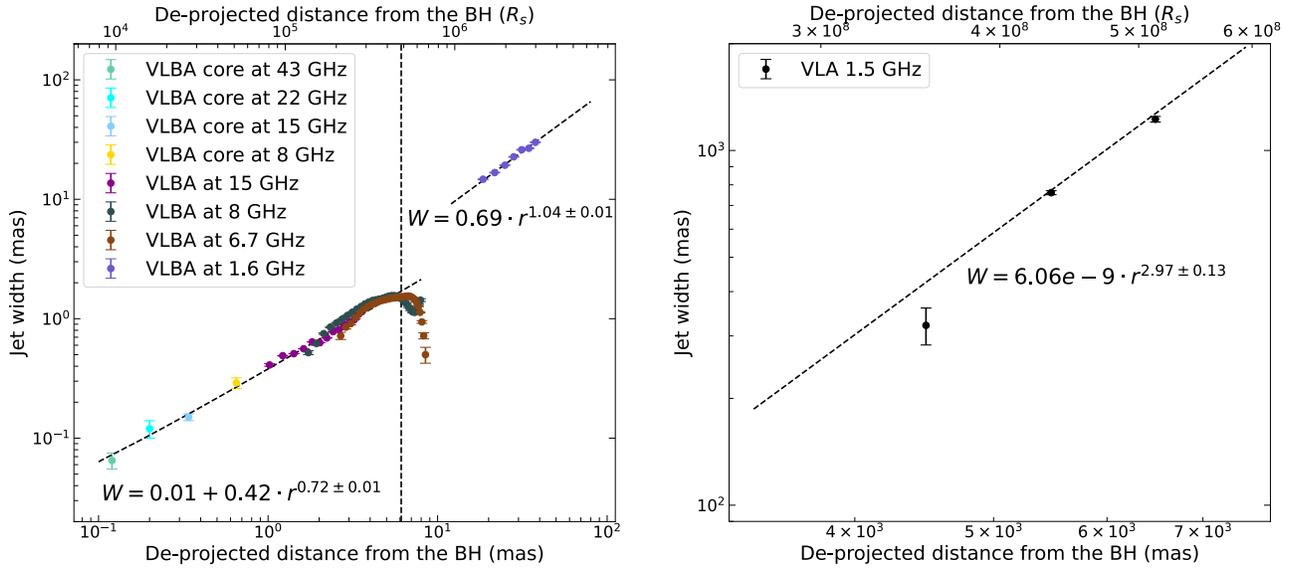

**Figure 11.** The jet width vs. radial distance. The dashed diagonal line represents the best-fit curve, and the vertical dashed line indicates $r = 6.1$ mas, where the jet begins to narrow in width.

2. We used five pairs of data from VLBA 1.6 to 43 GHz to determine the core shift of 0241+622. We performed least-squares fitting for the position of each frequency relative to the 43 GHz core using the formula $\Delta r_{\mathrm{core}} = a\nu^b + c$. The fitting results are $a = 8.10 \pm 1.27$, $b = -1.12 \pm 0.09$, and $c = -0.12 \pm 0.03$, which indicates that the jet base is located approximately 0.12 mas upstream of the 43 GHz core.

3. Based on the previously published model fits and the proper motion results of the jet components for 0241+622, we calculated the intrinsic Lorentz factor and the angle with the line of sight of the jet. The results show that the Lorentz factor is $\Gamma \geqslant 1.84$, and the angle between the jet and the line of sight is $\theta \leqslant 65°.7$. By using the brightness ratio of the jet and counterjet at 43 GHz for this source, along with the results from M. L. Lister et al. (2019), we obtained the velocity field of the source within 3.14 mas from the central black hole. We found that the source accelerates within this range, which is consistent with the jet shape we obtained.

4. We found that the jet width of 0241+622 begins to decrease at a distance of about 6.1 mas from the core, which may be a potential indication of a jet recollimation shock. We speculate that the decrease in jet width is caused by changes in external pressure at the Bondi radius. Based on the angle between the jet and the line of sight and the black hole mass, we calculated that the Bondi radius of 0241+622 is between 1.26 and 12.6 mas, which is consistent with the location where we observed the change in jet width.

5. In the low-frequency (VLA 1.5 GHz) images, we observed morphological changes in the jet of 0241+622 on larger scales. There is a bright jet component about 5″ from the core, but the jet structure in between is relatively weak. This may be because the external medium may become denser, and interactions between the jet and this medium generate shocks, locally dissipating kinetic energy and producing bright, compact emission points, corresponding to the observed bright jet components.

These findings provide new perspectives for understanding the jet structure and evolution process of 0241+622, as well as observational evidence for the mechanisms of jet collimation and recollimation. Future high-frequency and high-time-resolution VLBI observations will help further reveal the detailed characteristics and dynamic evolution of the jet.

### Acknowledgments

This work was supported by the National Key R&D Programme of China (2018YFA0404602). This research has made use of data from the MOJAVE database that is maintained by the MOJAVE team (M. L. Lister et al. 2018). The Very Long Baseline Array is operated by the National Radio Astronomy Observatory, a facility of the National Science Foundation, operated under cooperative agreement by Associated Universities, Inc. We used in our work the Astrogeo VLBI FITS image database, doi:10.25966/kyy8-yp57, maintained by Leonid Petrov.

## Appendix
## Stacking Images

To minimize the $\sigma_{\mathrm{rms}}$ in the final image as much as possible, we can stack multiple different observational data sets at the same frequency (e.g., B. Boccardi et al. 2016; A. B. Pushkarev et al. 2017; K. Akiyama et al. 2018; Y. Y. Kovalev et al. 2020; C. Casadio et al. 2021; X. Yan et al. 2023) or directly use data with a sufficiently long observation time (e.g., T. Haga et al. 2015; G. Giovannini et al. 2018; K. Hada et al. 2018; S. Nakahara et al. 2018; J. Park et al. 2021; H. Okino et al. 2022; K. Yi et al. 2024). Therefore, we stacked all the 6.7 GHz images together and performed the same operation on the 8 and VLA 1.5 GHz images, which allowed us to better recover the complete cross section of the jet (C. Casadio et al. 2021).

## ORCID iDs

Haitian Shang https://orcid.org/0009-0002-8909-2935
Wei Zhao https://orcid.org/0000-0003-4478-2887
Xiaoyu Hong https://orcid.org/0000-0002-1992-5260
Xu-zhi Hu https://orcid.org/0000-0002-5398-1303

## References

Akiyama, K., Asada, K., Fish, V. L., et al. 2018, Galax, 6, 15
Asada, K., & Nakamura, M. 2012, ApJL, 745, L28
Asada, K., Nakamura, M., Doi, A., Nagai, H., Inoue, M., et al. 2014, ApJL, 781, L2
Begelman, M. C., & Li, Z.-Y. 1994, ApJ, 426, 269
Beskin, V. S., Chernoglazov, A. V., Kiselev, A. M., & Nokhrina, E. E. 2017, MNRAS, 472, 3971
Beskin, V. S., & Nokhrina, E. E. 2006, MNRAS, 367, 375
Biretta, J. A., Sparks, W. B., & Macchetto, F. 1999, ApJ, 520, 621
Biretta, J. A., Zhou, F., & Owen, F. N. 1995, ApJ, 447, 582
Blandford, R. D., & Königl, A. 1979, ApJ, 232, 34
Boccardi, B., Krichbaum, T. P., Bach, U., et al. 2016, A&A, 585, A33
Boccardi, B., Krichbaum, T. P., Ros, E., & Zensus, J. A. 2017, A&ARv, 25, 4
Boccardi, B., Migliori, G., Grandi, P., et al. 2019, A&A, 627, A89
Boccardi, B., Perucho, M., Casadio, C., et al. 2021, A&A, 647, A67
Boccardi, G., Lillo, G., Mastrullo, R., et al. 2020, JPhCS, 1599, 012055
Boettcher, M., Harris, D. E., & Krawczynski, H. 2012, Relativistic Jets from Active Galactic Nuclei (Weinheim: Wiley)
Bogovalov, S., & Tsinganos, K. 2005, MNRAS, 357, 918
Britzen, S., Qian, S. J., Steffen, W., et al. 2017, A&A, 602, A29
Burd, P. R., Kadler, M., Mannheim, K., et al. 2022, A&A, 660, A1
Camenzind, M. 1987, A&A, 184, 341
Casadio, C., MacDonald, N. R., Boccardi, B., et al. 2021, A&A, 649, A153
Cheng, X. P., An, T., Frey, S., et al. 2020, ApJS, 247, 57
Cohen, M. H., Meier, D. L., Arshakian, T. G., et al. 2015, ApJ, 803, 3
Croke, S. M., & Gabuzda, D. C. 2008, MNRAS, 386, 619
Daly, R. A., & Marscher, A. P. 1988, ApJ, 334, 539
Eichler, D. 1993, ApJ, 419, 111
Fromm, C. M., Ros, E., Perucho, M., et al. 2013a, A&A, 557, A105
Fromm, C. M., Ros, E., Perucho, M., et al. 2013b, A&A, 551, A32
Giovannini, G., Savolainen, T., Orienti, M., et al. 2018, NatAs, 2, 472
Globus, N., & Levinson, A. 2016, MNRAS, 461, 2605
Greisen, E. W. 2003, in Information Handling in Astronomy—Historical Vistas, ed. A. Heck (Dordrecht: Kluwer), 109
Hada, K., Doi, A., Kino, M., et al. 2011, Natur, 477, 185
Hada, K., Doi, A., Wajima, K., et al. 2018, ApJ, 860, 141
Hada, K., Park, J. H., Kino, M., et al. 2017, PASJ, 69, 71
Haga, T., Doi, A., Murata, Y., et al. 2015, ApJ, 807, 15
Heckman, T. M., & Best, P. N. 2014, ARA&A, 52, 589
Homan, D. C., Cohen, M. H., Hovatta, T., et al. 2021, ApJ, 923, 67
Hovatta, T., Aller, M. F., Aller, H. D., et al. 2014, AJ, 147, 143
Hovatta, T., Valtaoja, E., Tornikoski, M., & Lähteenmäki, A. 2009, A&A, 494, 527
Hutchings, J. B., Campbell, B., Gower, A. C., Crampton, D., & Morris, S. C. 1982, ApJ, 262, 48
Komatsu, E., Smith, K. M., Dunkley, J., et al. 2011, ApJS, 192, 18
Komissarov, S. S. 2011, MmSAI, 82, 95
Komissarov, S. S., Barkov, M. V., Vlahakis, N., & Königl, A. 2007, MNRAS, 380, 51
Kovalev, Y. Y., Pushkarev, A. B., Nokhrina, E. E., et al. 2020, MNRAS, 495, 3576
Kutkin, A. M., Sokolovsky, K. V., Lisakov, M. M., et al. 2014, MNRAS, 437, 3396
Li, X., Mohan, P., An, T., et al. 2018, ApJ, 854, 17
Li, Z.-Y., Chiueh, T., & Begelman, M. C. 1992, ApJ, 394, 459
Liodakis, I., Marchili, N., Angelakis, E., et al. 2017, MNRAS, 466, 4625
Lister, M. L., Aller, M. F., Aller, H. D., et al. 2018, ApJS, 234, 12
Lister, M. L., Homan, D. C., Hovatta, T., et al. 2019, ApJ, 874, 43
Lyubarsky, Y. 2009, ApJ, 698, 1570
Marscher, A. P., Jorstad, S. G., D'Arcangelo, F. D., et al. 2008, Natur, 452, 966
McKinney, J. C., Tchekhovskoy, A., & Blandford, R. D. 2012, MNRAS, 423, 3083
Meier, D. L. 2012, Black Hole Astrophysics: The Engine Paradigm (Berlin: Springer)
Mertens, F., Lobanov, A. P., Walker, R. C., & Hardee, P. E. 2016, A&A, 595, A54
Meyer, E. T., Sparks, W. B., Biretta, J. A., et al. 2013, ApJL, 774, L21
Montgomery, D., & Runger, G. 2010, Applied Statistics and Probability for Engineers (New York: Wiley), https://books.google.com/books?id=_f4KrEcNAfEC
Nakahara, S., Doi, A., Murata, Y., et al. 2018, ApJ, 854, 148
Nakahara, S., Doi, A., Murata, Y., et al. 2019, ApJ, 878, 61
Nakahara, S., Doi, A., Murata, Y., et al. 2020, AJ, 159, 14
Nakamura, M., & Asada, K. 2013, ApJ, 775, 118
Nokhrina, E. E., Kovalev, Y. Y., & Pushkarev, A. B. 2020, MNRAS, 498, 2532
Okino, H., Akiyama, K., Asada, K., et al. 2022, ApJ, 940, 65
Park, J., Hada, K., Kino, M., et al. 2019, ApJ, 887, 147
Park, J., Hada, K., Nakamura, M., et al. 2021, ApJ, 909, 76
Perucho, M., Kovalev, Y. Y., Lobanov, A. P., Hardee, P. E., & Agudo, I. 2012, ApJ, 749, 55
Phillipson, R. A., Vogeley, M. S., & Boyd, P. T. 2023, MNRAS, 518, 4372
Pushkarev, A. B., Kovalev, Y. Y., Lister, M. L., & Savolainen, T. 2017, MNRAS, 468, 4992
Ro, H., Yi, K., Cui, Y., et al. 2023, Galax, 11, 33
Rybicki, G. B., & Lightman, A. P. 1979, Radiative Processes in Astrophysics (New York: Wiley)
Shepherd, M. C., Pearson, T. J., & Taylor, G. B. 1994, BAAS, 26, 987
Tchekhovskoy, A., McKinney, J. C., & Narayan, R. 2008, MNRAS, 388, 551
Traianou, E., Krichbaum, T. P., Boccardi, B., et al. 2020, A&A, 634, A112
Tseng, C.-Y., Asada, K., Nakamura, M., et al. 2016, ApJ, 833, 288
Véron-Cetty, M. P., & Véron, P. 2006, A&A, 455, 773
Vlahakis, N., & Königl, A. 2004, ApJ, 605, 656
Walker, R. C., Hardee, P. E., Davies, F. B., Ly, C., & Junor, W. 2018, ApJ, 855, 128
Yan, X., Lu, R.-S., Jiang, W., Krichbaum, T. P., & Shen, Z.-Q. 2023, ApJ, 957, 32
Yi, K., Park, J., Nakamura, M., Hada, K., & Trippe, S. 2024, A&A, 668, A94